\documentclass[compsoc,conference,a4paper,10pt,times]{IEEEtran}

\usepackage{cite}
\usepackage{amsmath,amssymb,amsfonts}
\usepackage{algorithmic}
\usepackage{graphicx}
\usepackage{textcomp}
\usepackage{bmpsize}
\usepackage{xcolor}
\usepackage{lipsum}
\usepackage[colorlinks=true,urlcolor=black]{hyperref}
\def\BibTeX{{\rm B\kern-.05em{\sc i\kern-.025em b}\kern-.08em
    T\kern-.1667em\lower.7ex\hbox{E}\kern-.125emX}}

\newif\ifanon\anonfalse
\newif\iffullversion\fullversionfalse 
\newif\ifcomments \commentsfalse 

\newcounter{casestudy}

\usepackage{latexsym}
\usepackage[colorlinks=true,urlcolor=black]{hyperref}
\usepackage{multirow}
\usepackage{bbm}
\usepackage{mathtools}
\usepackage{tikz}
\usepackage{tabularx}
\usepackage{color,soul}
\usepackage{float} 
\usepackage{xspace}
\usetikzlibrary{arrows.meta}
\usetikzlibrary{calc,shapes.callouts,shapes.arrows}
\usetikzlibrary{positioning,chains,fit,shapes,calc}

\usepackage[capitalise]{cleveref}

\crefname{thm}{Theorem}{Theorems}

\ifcomments
    \newcommand{\sunoo}[1]{[\textcolor{magenta}{Sunoo: #1}]}
    \newcommand{\daji}[1]{[\textcolor{blue}{Daji: #1}]}
    \newcommand{\elettra}[1]{[\textcolor{violet}{Elettra: #1}]}
\else
    \newcommand{\sunoo}[1]{\ignorespaces}
    \newcommand{\daji}[1]{\ignorespaces}
    \newcommand{\elettra}[1]{\ignorespaces}
\fi

\newcommand{\bi}[1]{\textbf{\textit{#1}}}

\newcommand{\subh}[1]{\smallskip \noindent \textbf{{#1}}.}
\renewcommand{\paragraph}[1]{\subh{#1}}

\newcounter{notec}

\newcommand{\note}[1]{%
  \refstepcounter{notec}
  \paragraph{\textit{Note \thenotec: #1}}%
}

    \usepackage[shortlabels]{enumitem}
    \setlist[itemize]{noitemsep, topsep=0pt, leftmargin=.2in}
    \setlist[enumerate]{noitemsep, topsep=0pt, leftmargin=.2in}

\newcommand{\qone}{How do interoperation and economic incentives clash
(or not)?\xspace}
\newcommand{\qtwo}{How do security arguments and economic incentives align (or not)?\xspace}
\newcommand{\qthree}{What needs to change in order for the company to support interoperation?\xspace}
\newcommand{\qfour}{What tension exists between security and other goals
both before and after?\xspace}

\newcommand{\question}[1]{\smallskip \noindent \textbf{{#1}}}
\begin{document}
\title{``Security vs. Interoperability'' Arguments: An Analytical Framework }

\author{\IEEEauthorblockN{Daji Landis}
\IEEEauthorblockA{\textit{New York University}\\
New York City, USA}
\and
\IEEEauthorblockN{Elettra Bietti}
\IEEEauthorblockA{\textit{Northeastern University}\\
Boston, USA} 
\and
    \IEEEauthorblockN{Sunoo Park}
\IEEEauthorblockA{\textit{New York University } \\
New York City, USA}
}

\maketitle

\begin{abstract}
Concerns about big tech's monopoly power have featured prominently in recent media and policy discourse, as regulators in the European Union (EU), the United States (US), and beyond have ramped up efforts to promote healthier market competition. One favored approach is to require certain kinds of \emph{interoperation} between platforms, to mitigate the current concentration of power in the biggest companies.
Unsurprisingly, interoperability initiatives have generally been met with resistance by big tech companies. Perhaps more surprisingly, a significant part of that pushback is in the name of \emph{security}---that is, arguing against interoperation on the basis that it will undermine security.

We conduct a systematic examination of ``security vs. interoperability'' (SvI) discourse in the context of EU antitrust and competition proceedings. Our resulting contributions are threefold. First, we propose a \emph{taxonomy} of SvI concerns in three categories: engineering, vetting, and hybrid. Second, we present an \emph{analytical framework} for assessing real-world SvI concerns, and illustrate its utility by analyzing several \emph{case studies} spanning our three taxonomy categories. Third, we undertake a \emph{comparative analysis} that highlights key considerations around the interplay of economic incentives, market power, and security across our diverse case study contexts, identifying common patterns in each taxonomy category. Our contributions provide valuable analytical tools for experts and non-experts alike to critically assess SvI discourse in today's fast-paced regulatory landscape.
\end{abstract}


\section{Introduction}

Billions of users entrust vast amounts of personal data to their devices and to services that are controlled by a small number of very large companies. These companies have huge discretion over with whom they share this data, and how they secure these vast data flows---in messaging, digital payments, tap-and-go technology, and more. 

Recent concerns around concentration of power in big tech have led to calls and legal mandates for platforms to support increased interoperation \cite{ll1, ll2, ll3, ll4, ll5}. The idea is that interoperation would open the market to more smaller players, promoting healthier competition, 
especially where a single market actor holds expansive control over a software and/or hardware ecosystem. 
Large tech platforms, on the other hand,  have every incentive to resist moves to open up their systems to interoperation. 

In this context, companies seeking to resist interoperability mandates are increasingly arguing that the security of their systems requires keeping them closed and limiting competition. They describe their systems' security as a key business proposition, and a key reason consumers are attracted to their business models.

 By way of illustration, Google recently argued against mandated interoperation with alternative app stores in a case in front of the US Supreme Court:

\begin{quote}
    ``[t]he [interoperation] requirement \dots effectively requir[es] Google to endorse stores that might be full of harmful content, ranging from malware that can scam or extort users to pornography and hate speech.''\footnote{Application for Partial Stay of Permanent Injunction Pending Disposition of Petition for a Writ of Certiorari at 39, \emph{Google LLC v. Epic Games, Inc.}, 147 F.4th 917 (9th Cir. July
31, 2025) (No. 25A), \emph{cert. denied}, No. 25-521 (2026).} 
\end{quote}

These kinds of statements tend to appear in public-facing communications and sometimes legal or regulatory filings. As such, they typically remain high level, and do not get into technical detail on the precise nature of the claimed tensions between security and interoperation.

Such ``security vs. interoperability'' arguments are often framed as technical arguments about incurring unacceptable security risks---an area outside the usual expertise of courts and policymakers, and squarely within that of the tech companies subject to regulation.\footnote{Not only because of technological expertise in general, but also due to companies' detailed knowledge of how their own systems work.}
Perhaps the most prominent past instance where companies made technical arguments about incurring unacceptable security risks was in resisting government efforts to weaken encryption \cite{fbi_case}---in that context, the companies were widely seen as the champions of the public interest.\footnote{These security-critical arguments may well have served the companies’ business interests as well.}
Arguments about the security risks of interoperation may carry particular weight---for policymakers, regulators, and the general public---when raised by the same companies,  which are already responsible for vast amounts of sensitive data that could be imperiled by ill-advised regulation and have established reputations for security and privacy.

Although companies should be able to put forward the best case for their own interests in such regulatory contexts, this trend is concerning if and when seemingly technical arguments are treated with more deference than business or policy arguments, as more objective or factual, or as matters on which courts and regulators should defer to companies' technological expertise.

The defining characteristic of ``security vs. interoperability'' arguments is the potential for them to instrumentalize ``security'' into a cause for deference that is convenient for business ends.
Consider the quote above---Google is right: alternative app stores may contain lots of malware. \emph{But does this argument mean that regulators should drop their antitrust efforts because of the grave security harm that Google argues will result?} That does not follow directly. The security community might say that is a question of antitrust law or policy and beyond our domain---but the answer to that antitrust question hinges on a security question beyond antitrust experts' domain. 

The ``security vs. interoperability'' narrative is an inherently interdisciplinary one, about acceptable tradeoffs in integrating technical security goals with other policy goals. 
The availability of alternative app stores raises security risks, which may be balanced against other benefits, such as a more competitive marketplace.
Focusing on only technical aspects or only economic or regulatory aspects yields an inherently incomplete picture, and policymaking based on such a view could lead to missteps in the important and fast-moving area of big tech and antitrust. 
Thus, it is critical for the security community to
engage with ``security vs. interoperability'' questions 
not by opining on isolated technical questions
but by embracing the broader economic and policy context in which these arguments appear---especially when industry interests may not align with highlighting this broader scope.
\iffullversion\else
Our engagement is essential to complement lawyers, regulators, and policymakers in this impactful space, to clarify and mitigate potential security harms of antitrust initiatives.
\fi

This paper provides a blueprint for doing just that, based on a systematic review of relevant cases in the EU. We present a \emph{taxonomy} that divides ``security vs. interoperability'' tensions into three key categories: engineering, vetting, and hybrid. The hybrid case combines aspects of both engineering and vetting issues but exhibits more complex dynamics than either alone. Our \emph{analytical framework}, illustrated via a series of detailed \emph{case studies} from the EU, captures the key economic considerations relevant to the regulatory context in which the ``security vs. interoperability'' discourse is rapidly unfolding.\footnote{The EU has focused heavily on Apple in its recent interoperation-related competition efforts. As a result, our case studies mostly involve Apple, as further discussed in Section~\ref{sec:method} (Note \ref{note:apple}). 
} Our framework is designed to be usable without knowledge of antitrust law or economics, and captures key aspects and patterns across all of the real-world cases we systematically examine. 
For each case study, we provide detailed analysis of the security, interoperability, and incentives considerations.
These case study analyses illustrate our approach's utility in disentangling where security and interoperability are and are not in tension, where securing interoperable systems presents novel engineering challenges, and 
where ``security arguments'' against interoperability may be a distraction from the real issue: improved competition.
Finally, our \emph{comparative analysis} identifies common patterns in the interplay of incentives, market power, and security in each taxonomy category.

\smallskip
\paragraph{Broader context and motivation}
The use and misuse of security arguments is an issue of natural interest to the security community.
In the present context, the broader implications lie as much in security outcomes as in competition policy outcomes, and the timeliness and potential breadth of impact of the ``security vs. interoperability’’ issue arises more from today’s competition policy context than from the field of security.

The concentration of economic and political power in the hands of a few technology companies, alongside the novelty, complexity, and opacity of modern technology, has contributed to a climate in which companies may strategically deploy (seemingly) technical arguments to further their political and economic power and avoid regulation. Recent legal scholarship has highlighted companies' strategic use of 
privacy, content moderation, and security arguments to further economic and policy interests \cite{privacy_pretext,lochner_law_review, fidler_cybersecurity_2023,cohen_book, san_bern_law_review}, including specifically in the context of the EU competition enforcement~\cite{brown2022messaging, brown2025iosinterop}. Van Loo~\cite{privacy_pretext} discusses how companies use privacy inconsistently as a pretext to further business aims, including to limit competitors' access to data, while engaging widely in data selling and sharing when beneficial for business.

This paper focuses on companies' security-related discourse in the context of competition investigations in the European Union that involve mandated interoperability.
When companies raise security arguments in this context, in general, the criticality and nuances of the security issue cannot be expected to be transparent to lawyers, courts and policymakers, or to other public stakeholders. 
 
Companies seeking to defend themselves from competition claims may strategically bring in security experts. Those seeking to impose competition obligations will, in turn, bring their own security experts. In the \emph{Epic v. Apple} case, companies' experts have asserted that a system might suffer serious security risks or degradation if the company is forced to comply with interoperability obligations, and the opposing experts have argued that this is not true~\cite{epic_apple_blog}.\footnote{The ``battle of the experts'' problem is a much broader problem with expert witnesses in legal systems \cite{experts}.}

We believe our systematic analysis of ``security vs. interoperability'' discourse can help the security and policy communities better analyze and conceptualize “security vs. interoperability” tensions (and the lack thereof), leading to better policy decisions that will shape the future of technology markets and ecosystems. 

Those policy decisions may go either way, depending on the context---we are not arguing for particular outcomes. We caution against ``security vs. interoperability'' arguments' potential use as a pretext to further business aims; equally, we caution against underestimating the real security challenges in building interoperable systems. These two cautions go hand in hand: distinguishing one from the other is the crux of the difficulty for judges and policymakers, especially when business interests favor companies making both these kinds of arguments and making them hard to distinguish from one another.

\paragraph{Other related work}
That security engineering requires consideration of contextual sociotechnical factors is hardly a new observation~\cite{ross_book}. The security economics literature has long highlighted and studied the idea that ``security failure is caused by bad
incentives at least as often as by bad design'' \cite{AM07,ross_econ}. Much of security economics considers how incentives influence security outcomes. Our inquiry takes a different angle, in a similar spirit: we ask how security arguments can be leveraged to serve other economic goals.

Digital competition policy and antitrust enforcement have been responding to trends in digital markets long before the recent wave of interest in big tech and antitrust. Microsoft's bundling of Internet Explorer with Windows was the subject of a landmark 1998 antitrust case.\footnote{United States v. Microsoft Corp., 253 F.3d 34 (D.C. Cir. 2001).} Net neutrality regulations, which required internet service providers not to give preferred treatment to certain types of content or traffic for economic reasons, have been at the heart of political controversies since the 1990s \cite{nyt_net_neutrality, nn1, nn2, nn3, nn4, nn5, nn6}. The ``security vs. interoperability'' narrative, our focus in this paper, is a more recent development that very little prior work has considered. Since the European Union's Digital Markets Act (DMA), a promising initial literature has begun to examine the technical complexities of securely  interoperating end-to-end encrypted messaging~\cite{dma_concerns_e2ee_ross,dma_how_to_e2ee}.  

Other prior work, which we discuss where relevant throughout the paper, include work on the difficulty of interoperation in specific contexts \cite{PGP, Halpin20, KRAPVC18, LRSS13, MLS, eth-hf,PSNR21}; tension between security and usability \cite{usability_book}; and treatments in the security literature of the specific technologies we address \cite{Darmstadt_18, Darmstadt_19, Darmstadt_21, BLE_not_private, ble_not_private_2, apple_bluetooth, imessage_v_signal, browser_sok, browser_security, evil, clark_email, bluetooth_mitm, pix_fraud, superapp_sok}.  We also draw from the economics literature \cite{GARCIASWARTZ2019110, SHN21, hor_vert_interop, ll1, evans2011platform} and law and policy literature \cite{ll5, cohen_book, imessage_doj, experts, ll4, privacy_pretext, san_bern_law_review, lochner_law_review, ll3, fidler_cybersecurity_2023,ll2, cohen2017platforms, brown2022messaging, brown2025iosinterop}.

\smallskip
\paragraph{Summary of contributions}
We critically examine ``security vs. interoperability'' (SvI) discourse related to competition proceedings in the EU.
Our four key contributions are as follows.

\begin{enumerate}
    \item We conduct a \textbf{systematic review of EU competition cases} involving interoperability mandates for digital technology that feature SvI considerations   (\S \ref{sec:method}).
    \item We propose a three-part \textbf{taxonomy} of SvI concerns, and explain the taxonomy's
    relationship to the economic concepts of \emph{vertical} and \emph{horizontal} interoperation  (\S \ref{sec:framework}). 
    \item We introduce an \textbf{analytical framework} centered around four critical questions, and apply it to \textbf{six case studies} across diverse contexts featuring SvI concerns: messaging, app sideloading, 
    app stores, browser engines, NFC,
    and physical device interoperation (\S\ref{sec:case-studies}).
    \item We offer a \textbf{comparative analysis} of these case studies through the lens of our framework, illuminating \emph{key patterns in the interplay of economic incentives, market power, and security} in each taxonomy category
    (\S\S\ref{sec:case-studies}--\ref{sec:big-picture}).
\end{enumerate}

\section{Our Methodology}
\label{sec:method}
We systematically reviewed cases where security comes up in the context of mandated interoperability in competition proceedings. We focus on EU cases, as there has recently been an extensive and well documented surge of relevant competition enforcement activity in the EU. Similar issues have arisen in US antitrust proceedings; we will note examples of prominent parallel US cases.  We also briefly discuss a few key cases outside the EU in Section \ref{sec:generalizability}.

We first collected all cases in which there have been interoperability mandates related to digital technology, in the EU Commission's competition case database~\cite{ec_database} (henceforth, `EU Database'). 
We then searched for discussions about security in the context of our selected cases.  The resulting collection of discourse---from legal and regulatory proceedings, press releases and press coverage, developer-facing documentation, and consumer-facing information---forms the basis for our systematization and analysis.  

The rest of this section details: (1) how we selected our cases; (2) high-level results of our screening process;
(3) how we tagged these cases with relevant technologies; and (4) how we identified relevant security issues.

\paragraph{What we mean by `interoperation'} We use \emph{interoperation} to refer broadly to practices in which products or services produced by one company (e.g., smartphones) support or facilitate user engagement with products or services developed by a third-party (e.g., independent apps).
For our purposes, interoperation comprises both \emph{technical interoperation} (e.g., what are the technical requirements for an app to run on a given platform?) and \emph{interoperability policy} (e.g., which technically interoperable apps does a platform permit as a matter of policy?). We write \emph{interoperability} to mean \emph{the ability to interoperate}. 

\subsection{Search Methodology}

The EU Database features five categories of `policy areas,' of which only three are within our scope: `Antitrust \& Cartels,' `Digital Markets Act,' and `Merger.'\footnote{The other categories are `Foreign Subsidies' and `State Aid.'}

In our initial search, we searched all cases in either the `Antitrust \& Cartels' or the `Digital Markets Act' policy area. 
In a follow-up search, we covered `Mergers' with narrowed search criteria (including limited economic activities), a refinement we discuss below.  
In both the initial and the follow-up search, we used the search terms `interoperate interoperability interoperation,' which the EU Database search engine interprets as a disjunction. 

Given the specific definition of interoperation we seek to capture, we conducted a preliminary search of DMA cases in the EU Database to check whether our search terms were used with the meaning we intended to capture. We found that the search terms were used in reference to both \emph{technical interoperability} and \emph{interoperability policy}, as defined above. We also tried other similar terms such as `compatible' or `integrated' and found these did not capture any cases we deemed relevant that were not in our original search. 

From the results of our initial and follow-up searches, we manually screened for cases satisfying the following additional criteria: (1) the case concerned interoperation of digital technologies (not, e.g., trains), and (2) the case involved interoperation mandated by competition authorities.
Section~\ref{sec:manual} offers more detail on the screening process and as does a Figure \ref{fig:case_selection_flowchart} in the appendix.

Based on our initial search of  Antitrust and DMA cases, we manually identified which classifications of economic activities\footnote{These are classifications of industry made by the EU \cite{nace_codes}} included relevant digital technologies.
We used these classifications\footnote{We included these codes: NACE Rev 2 (2008): C.26.1, C.26.2, C.26.3, C.26.4, C.26.51, J.61.2, J.61.9, J.62, J.63, K.64.99; NACE Rev 2.1 (2025): C.26.1, C.26.2, C.26.3, C.26.4, C.26.51, K, L.64.9.} to filter and improve efficiency in our follow-up search of Merger cases.\footnote{We first eliminated codes that were unlikely to be about digital technology (e.g., 
C.10.52 - Manufacture of ice cream), and included any codes directly pertaining to digital technology. For the remaining codes (e.g., regarding finance), we omitted codes associated only with one or more non-relevant cases in the initial search, and kept all others.} 

\subsection{Screening for Relevant Cases}
\label{sec:manual}

Our screening process consisted of two main phases: an initial screening based on a quick read, and a second phase where a more thorough reading was necessary.

\paragraph{Phase 1 (P1)} We filtered out any cases that were dismissed or that were just DMA `gatekeeper' designations\footnote{\label{fn:gatekeeper}The term gatekeeper as defined in the DMA Article 3 refers to entrenched incumbent companies that provide platforms that are important for third parties to be able to reach end users.} as these cannot directly give rise to interoperability mandates.
We moreover filtered out cases that were quickly identifiable as out of scope, for one of the following reasons. First, many cases used the term interoperability in the text of a common definition of `Standards'---we discarded cases where this was the only mention of the term. Second, some cases mentioned interoperability a few times only as an existing product feature or part of the regulatory context; these cases did not involve mandates in their mentions of interoperability, so we discarded them. Third, sometimes multiple cases appear in the same document; we discarded cases in the same document as a mention of interoperability that do not themselves mention the term.  
Phase 1 eliminated the majority of irrelevant cases (see Figure \ref{fig:case_selection_flowchart}).

\begin{figure}[htbp]
    \centering
    \begin{tikzpicture}[
        box/.style={rectangle, draw=black, thick, align=center, minimum width=5.5cm, inner sep=3mm, font=\small},
        exbox/.style={rectangle, draw=black, thick, align=left, minimum width=3.8cm, inner sep=3mm, font=\small},
        arrow/.style={->, >=Stealth, thick}
    ]

    \node (initial) [box] {\textbf{Initial Cases Identified}\\ $n = 111$\\ (Antitrust: 28, DMA: 17, Merger: 66)};
    
    \node (remaining) [box, below=3cm of initial] {\textbf{Cases Remaining after P1}\\ $n = 31$\\ (Antitrust: 6, DMA: 4, Merger: 21)};
    
    \node (final) [box, below=3cm of remaining] {\textbf{Final Cases Kept}\\ $n = 14$\\ (Antitrust: 2, DMA: 4, Merger: 8)};

    \path (initial.south) -- (remaining.north) coordinate[midway] (mid1);
    \path (remaining.south) -- (final.north) coordinate[midway] (mid2);

    \node (ex1) [exbox, right=1.2cm of mid1] {\textbf{Excluded in P1}\\ $n = 80$\\ \textbullet\ Antitrust: 22\\ \textbullet\ DMA: 13\\ \textbullet\ Merger: 45};
    
    \node (ex2) [exbox, right=1.2cm of mid2] {\textbf{Excluded in P2}\\ $n = 17$\\ \textbullet\ Antitrust: 4\\ \textbullet\ DMA: 0\\ \textbullet\ Merger: 13};

    \draw [arrow] (initial) -- (remaining);
    \draw [arrow] (remaining) -- (final);

    \draw [arrow] (mid1) -- (ex1.west);
    \draw [arrow] (mid2) -- (ex2.west);

    \end{tikzpicture}
    \caption{Case selection and filtering process.}
    \label{fig:case_selection_flowchart}
\end{figure}

\paragraph{Phase 2 (P2)}   All cases that remained required a more thorough reading to establish whether they were within our scoping criteria. We removed a number of cases that discussed interoperability as a product feature or important context in greater depth than was readily apparent, but did not mention interoperability 
\emph{mandates}. We also removed one case that was not about digital technology. The resulting 14 cases are listed in Table~\ref{tab:all_cases} (Appendix).

\note{Merger case selection} \textit{Among the Merger cases, many discussed potential interoperability issues, often the degradation of existing interoperability, but did not ultimately have an interoperability mandate.  Generally this was because the Commission decided that the potential benefit to the merged entity of degrading interoperability was either not incentive compatible (i.e., not a good business decision) or not feasible. In a few cases the degrading was not deemed to be significant enough to greatly impact the market. In those cases we kept, there was some commitment regarding interoperation that was required for the merger to be allowed.}

\subsection{Identifying Relevant Technologies}
\label{sec:identify_technologies}
With these fourteen cases in hand, we next identified the technology or technologies that were the subject of (potential) interoperability mandates. The cases often discussed many products and not all were necessarily related to interoperability mandates. Some cases discussed multiple distinct technologies subject to interoperability mandates. Some distinct cases discussed the same type of technology.  We define the \emph{relevant technologies}, within a case context, to be the technologies subject to interoperability mandates. Table~\ref{tab:all_cases} (Appendix) lists relevant technologies for each case.

In some instances, a case we selected mentioned another, apparently similar case about similar technologies. This led to detailed review of one EU case beyond the 14 selected (on WhatsApp and Messenger).\footnote{Case DMA.100024 - Meta, number-independent interpersonal communications services, (2023)  OJ C, C/2023/1092, 05.09.2023.}

\subsection{Identifying SvI Considerations}
\label{sec:svi}

Next, we collected discussion about the security implications of interoperability mandates---that is, SvI considerations---in each of our selected cases, across three sources: (1) the relevant company’s public-facing statements about the case, (2) mentions of security in relevant legal and regulatory proceedings, and (3) developer-facing information regarding the technical details of the relevant interoperation.   
By SvI considerations, we mean any actual, potential, perceived, or claimed tensions between security goals and mandated interoperation.

We collected such discussion by searching the press releases and developer documentation of the companies involved during the relevant time period, as well as Google searches for the company name and relevant technology alongside various combinations of the search terms `merger,' `antitrust,' `DMA,' `EU,' `Commission,' `security,' and `privacy.'  Note that this process might not have captured all company communications, especially for older cases, as search interfaces and archives for press releases and developer documentation were variable.\footnote{Our oldest case was decided in 2004, and the oldest relevant press release we were able to find is now only available on the Internet Archive.}  We found company publications or news articles with official statements about every case, although many did not mention security or mandated interoperability. 

For mergers in particular, there was generally no discussion of the details of the commitments in public-facing venues.  The only exception to this was the Google/Fitbit merger.  As mergers are voluntary, they tend to involve situations in which companies are incentivized to make the merger appear favorable to shareholders and customers. In contrast to other competition proceedings, therefore, companies may be disincentivized from publicly suggesting that a merger will be disadvantageous to their security.\footnote{Companies may make security arguments to the Commission that are not public \cite{law_book}, to which we did not have access.}

Not every selected case involved an SvI concern. For some cases, we did not find any mentions of security at all: for example, the Cisco/Tandberg merger did not discuss the security of the underlying technology at all.\footnote{Some of the mergers had surrounding discussion that said the merger and resulting interoperability (independent of any mandates) would \emph{improve} consumer security as in the Intel/McAfee case.}

\note{Interoperability without security} \textit{
For those cases where we found no relevant security issues, we found a natural sub-categorization emerged, according to the type of interoperability mandate.  All such cases we examined either required (1) that interoperability not be degraded in future, (2) that information related to interoperability be disclosed (but no further facilitation of interoperability was required), or (3) both. Table~\ref{tab:all_cases} (Appendix) marks such cases as having `No' SvI concern, followed by a sub-category (`Degrade,' `Disclose,' or `Both').}

\note{Prevalence of Apple cases}\label{note:apple} \textit{The EU has initiated many proceedings against Apple in particular.  In the cases pertaining to app stores (Section \ref{sec:case-studies:vetting}), the EU has focused on Apple alone. In the US, cases brought by Epic Games, maker of Fortnite, against \emph{both} Apple and Google, alleged similar antitrust behavior from both companies.\footnote{\emph{See} Epic Games, Inc. v. Google LLC, 147 F.4th 917 (9th Cir. July 31, 2025); \emph{see also} Epic Games, Inc. v. Apple Inc., 67 F.4th 946 (9th Cir. Apr. 24, 2023).} Our case studies focus mostly on Apple, mirroring the cases captured in our EU-focused methodology. We believe the analogous behavior on the part of Google is sufficiently similar that our analysis is likely to generalize.}

\subsection{Systematization and Analysis}

Our detailed review of these collected cases and materials, found three clear patterns---or categories---in the nature of the security arguments and the nature of changes needed to comply with the relevant interoperability mandate and these form the basis for our taxonomy of SvI concerns (\S \ref{sec:framework}). 
The interplay of security and economic considerations that we observed repeatedly in the cases and materials formed the basis for our analysis questions (\S \ref{sec:case-studies}). 
We chose one to three case studies to feature for each taxonomy category, aiming to capture diverse and well-documented issues (\S \ref{sec:case-studies}).

\subsection{Limitations}
Our search methodology only covered the EU. While we make note of relevant US cases, we do not perform a systematic analysis of them.  
We are aware of other relevant cases in other jurisdictions that include both different regulatory strategies applied against the same companies and cases against companies we do not mention here. We will discuss generalizability in Section \ref{sec:generalizability}. We relied on keyword search, which is imperfect.  In the search for relevant security arguments, we may have missed examples that did not show up on Google, especially from the older cases which we found were more sparsely documented.

\section{Our Taxonomy}
\label{sec:framework}
\begin{table*}[ht]
\centering
\renewcommand{\arraystretch}{1.5} 
\begin{tabularx}{\textwidth}{|p{0.15\textwidth}|p{0.23\textwidth}|X|X|}

\hline
\textbf{Type of concern} & 
\textbf{Starting point} &
\textbf{Relevant security status quo} &
\textbf{Mitigation}  \\
\hline

\bi{Engineering} &
No \emph{technical} interoperability &
Technical security guarantees of \newline existing system design & Build something new (if possible; may require \emph{technical innovation})  \\ 
\hline
\bi{Vetting} &
No/onerous/costly interoperation as a matter of \emph{policy} &
Security protections of existing \newline interoperation policy & Vet third parties and products as a \newline condition of interoperation \\
\hline
\bi{Hybrid} & Integrated in-house interoperation, with no support for third parties & Security of in-house products and in-house product integration & Build technical support for third-party products and vet them \\
\hline
\end{tabularx}

\caption{Overview of Our Taxonomy}
\label{tab:security_arguments}
\end{table*}

\label{sec:framework:taxonomy}
We now introduce our three-part taxonomy, which distinguishes two broad types of ``security vs. interoperability'' arguments as raising either security \emph{engineering} concerns 
or security \emph{vetting} concerns, 
alongside a third \emph{hybrid} category that raises concerns incorporating aspects of both other categories. 
Table~\ref{tab:security_arguments} provides an overview.
Our taxonomy clarifies the different types of security concerns that may arise in response to the prospect of introducing interoperation between products or systems which did not previously interoperate. 

The technical, economic, and policy contexts in which these concerns are raised is critical to our taxonomy and subsequent analysis (\S\S \ref{sec:case-studies}--\ref{sec:big-picture}). In particular, the status quo of how the systems in question work \emph{before} interoperation is introduced, the incentives surrounding interoperation, the nature of the technical security guarantees the systems already provide, and the security policies and tradeoffs already in place, are essential considerations.

\paragraph{Security Engineering Concerns}
\label{sec:framework:taxonomy:eng}
When considering introducing interoperation to systems that are not technically interoperable, preserving security alongside interoperation may require that some aspect of the company's product or architecture be non-trivially changed. How to change systems to preserve the security guarantees currently offered, while also supporting interoperation, may or may not be clear based on existing techniques.

In such contexts, companies may raise concerns about the costs and engineering challenges that would be required to preserve security while supporting interoperability.
We call these \emph{security engineering concerns}.
\begin{center}
    \fbox{\parbox{8cm}{
    \textbf{Template of a Security \emph{Engineering} Concern} 
    \\ From a \emph{platform}'s perspective
    \begin{enumerate}
        \item Interoperating would require a building new technical mechanism or architecture for our system to interface with other systems.
        \item Our system has existing technical security guarantees. Designing a new architecture to preserve these guarantees is an engineering task that is costly in time, effort, and coordination with others---and may require innovation, so success is not certain.
        \item Therefore, interoperating secure systems is more costly than interoperating systems in general, and if we don't get the engineering right, the security of billions of users could be undermined.
    \end{enumerate}
    }}
\end{center}

Security engineering concerns tend to arise when there are two (or more) parties trying to interoperate similar services while maintaining the technical security guarantees of their existing system designs. Historically, in such situations, all parties involved need to develop, agree on, and implement compatible protocols or systems, so the security engineering concern is shared among all the parties seeking to interoperate. 
In contrast, the recent EU interoperation mandates allow `gatekeepers'\footnote{See footnote~\ref{fn:gatekeeper}.} to decide many aspects of the terms on which others may interoperate with them.
Messaging services provide a clear example of a security engineering concern (see Section~\ref{sec:case-studies:engineering}).

\note{Securely engineering interoperation is hard}\label{note:hard} \textit{ Coordinating between systems managed by different parties is inherently complex. Security engineering in such decentralized or federated contexts can present novel challenges not present in the context of centralized systems---as is well known in the security literature through a long history of examples such as PGP \cite{PGP, Halpin20}, TLS \cite{KRAPVC18}, DNSSEC \cite{LRSS13}, MLS \cite{MLS}, and blockchains \cite{eth-hf,PSNR21}. Each of these required a negotiated standard and widely distributed implementation and maintenance. Related challenges include consensus building and communication as well as consistent implementation and enforcement of security standards over time. It seems possible that the EU approach---of assigning broad authority to `gatekeeper'
entities to set the terms of interoperating with their systems---may alleviate some of these complications, especially as regards consensus building. That said, engineering interoperation with powerful `gatekeepers' still raises complex technical challenges and open questions (see our case study on messaging in Section~\ref{sec:case-studies:engineering}). }

\paragraph{Security Vetting Concerns}
\label{sec:framework:taxonomy:vet}
Security vetting concerns arise in contexts where there is already some kind of interoperation supported by a system's technical architecture. Many devices and platforms support an ecosystem of apps, software, and connected devices that are integral parts of their functionality, but that are developed by third parties. Examples include platforms, browsers, or operating systems on which third-party software can run, and platforms that offer API access to third-party developers.

Given such an architecture, there remains the question of which third parties a platform is willing to interoperate with.
Downloads and other connections can open a vector of attack for malware, spyware, and other undesirable outcomes for both the user and the device. Companies may vet and control which apps, systems, or software are permitted to interact with their devices with a view to preventing or mitigating such undesirable security outcomes---although these decisions might be motivated by other goals too.
\begin{center}
    \fbox{\parbox{8cm}{
    \textbf{Template of a Security \emph{Vetting} Concern} 
    \\ From a \emph{platform}'s perspective
    \begin{enumerate}
        \item Interoperating would mean our users and systems interacting with third parties and third-party software.
        \item Third parties and third-party software may make different decisions about security than us. We cannot guarantee that third-party software would adhere to the same standards as software we have developed or carefully examined ourselves. 
        \item Therefore, allowing third-party software on our platform could undermine billions of users' security. 
    \end{enumerate}
    }}
\end{center}

In such contexts, when asked to support greater interoperation, companies may raise concerns that third-party software that will interoperate with their systems may not be as secure as their own systems, and thus undermine the security of their users and/or systems. We call these \emph{security vetting concerns}.

Vetting third parties can be an important way to protect users, but it also concentrates power and discretion of incumbent companies over ecosystems they control.  Companies in this position can use their role as security guarantor to charge developers for access or prevent competitors from reaching the market. Platforms' interoperation policy choices thus require striking a tradeoff between security, interoperation, and other business considerations. 

Due to the discretionary tradeoffs involved, arguments that security considerations \emph{necessitate} a particular form of vetting or security measures generally cannot be substantiated by purely technical reasoning. Rather, they must refer to existing interoperation policies and the discretionary choices already present therein. This contrasts with security engineering concerns, which technical reasoning may suffice to substantiate, at least in some cases.

We see this with a few familiar examples: phones, personal computers (PCs),\footnote{Throughout, we use ``PC'' to refer personal computers. We do \emph{not} use the narrower definition of PC as a personal computer compatible with the original IBM PC. In particular, we consider Macs to be PCs.\label{fn:pc}} and video games. One clear strategy---arguably the only foolproof one---that would achieve the aim of protecting users from harms from third-party software would be to prevent phones from being able to download any software, and perhaps even bar them from the internet.  Such an approach would, of course, undermine the functionality of the product and be a bad economic move. Existing smartphone providers strike different tradeoffs; none take an absolute approach.

Whereas PC operating systems (OSs)  allow users to download almost anything, video-game consoles have much more stringent control over what is allowed on the machine.  This is not because of fundamental security differences between PCs and Xboxes, but because the respective developers 
have chosen different tradeoffs between security, interoperability, and other interests. 

PCs are key infrastructure for vast swaths of the economy and valued for versatile multi-purpose use. 
Exacting controls enacted by their manufacturers could make the devices undesirable and therefore not competitive in healthy market conditions, or, under monopoly conditions, deeply harm both users and the software industry. This is all the more so given historically permissive norms around what software can run on PCs.

In Section~\ref{sec:case-studies:vetting}, we provide three case studies of security vetting concerns: app sideloading, alternative app stores, and browser engines (WebKit).

\paragraph{Hybrid Concerns}
\label{sec:framework:taxonomy:hyb}
Finally, there are situations featuring aspects of both engineering concerns and vetting concerns, in which interoperation entails both the development of a secure system for interoperation and some sort of vetting of those developers looking to use that newly open system.
In hybrid cases, third parties must still demonstrate that their products meet the requirements of an interoperation policy; at the same time, there is also engineering and collaboration needed to support technical interoperability. 

Perhaps surprisingly, by combining engineering and vetting considerations, hybrid cases often exhibit more complex dynamics than either alone, due to their interplay. 
The same powerful company plays two roles, platform gatekeeper in a vetting context, and necessary collaborator in an engineering context. Third party developers must liaise with this same company for both.
For example, in opening mobile wallet functionality to third parties, device/OS providers need to both add APIs to allow developers access to previously restricted hardware and vet which developers will have access. Section~\ref{sec:case-studies:hybrid} offers two case studies of hybrid concerns: NFC for payments and physical device interoperability.
\begin{center}
    \fbox{\parbox{8cm}{
    \textbf{Template of a \emph{Hybrid} Concern} 
    \\ From a \emph{platform}'s perspective
    \begin{enumerate}
        \item Interoperating would mean allowing third parties to access functionality reserved for only our use.
        \item Opening those functionalities means third parties newly interacting with security-sensitive OS parts. Designing new systems to secure such access is a costly engineering task and may require innovation.
        \item Therefore, third-party access to these functionalities carries inherent security risk, and could undermine the security of billions of users. 
    \end{enumerate}
    }}
\end{center}

The combination of these two roles in one entity expands its leverage beyond what might be expected from simply ``interpolating'' engineering and vetting concerns, rendering hybrid cases arguably the most interesting and complex of the three types.
We discuss these relationships further, from an economic perspective, in Section~\ref{sec:framework:horiz-vertical}.

\paragraph{Summary}
Our Templates are designed to capture just the essential elements of the argument patterns we have identified. Of course, they are oversimplifications with respect to concerns being raised in any real-world context. Unsurprisingly, the arguments made in practice are more complex, and may not neatly fall into a single category.

Our case studies and analysis in Sections~\ref{sec:case-studies}--\ref{sec:big-picture} demonstrate the complexities of ``security vs. interoperability'' arguments in the wild, and show how recognizing the simple patterns identified in our taxonomy illuminates broader dynamics that can inform critical evaluation of the security and economic aspects of real-world arguments in specific case study contexts.

\subsection{Horizontal and Vertical Interoperation}
\label{sec:framework:horiz-vertical}

Our taxonomy is closely correlated to distinctions in the relationships of the interoperating products---\emph{horizontal} and \emph{vertical} relationships---corresponding to similar terminology in economics and competition law. 

To highlight this distinction, we define horizontal and vertical interoperation, generally in keeping with the use of these terms in \cite{hor_vert_interop}.  In a nutshell, by \emph{horizontal interoperation} we mean interoperation between products offered at the same level, 
i.e., products that run \emph{with} each other, 
whereas \emph{vertical interoperation} is that between a platform or `higher layer'\footnote{We use `layers' in a general sense, not specifically those from the OSI model, but generally to mean products that run on other products.} product and products that run \emph{on} that platform or higher layer product. 

Horizontal interoperation is when similar products need to be made interoperable. Messaging is a good example, where previously competing products must collaborate on integrated functionality and security (Case Study \ref{sec:case-studies:messaging}). This happens when the products are \emph{substitutable}, an economics term meaning two products used for similar purposes that compete for the same user demand \cite{econ_textbook}. 

Vertical interoperation happens between products on different layers. Instances of vertical interoperation that arise naturally (as opposed to through regulatory or antitrust enforcement) tend to be between some platform and a \emph{complementary} product. Again using economic terminology, we say two products are \emph{complementary} if they are often used together, or, formally, if a price increase of one decreases the demand for the other\cite{econ_textbook}. Smartphone apps are a good example, and the success of smartphones is greatly dependent on the wide availability of apps and other complementary products developed by third parties.

There is a clear incentive to achieve good vertical interoperation between platforms and the apps that run on them, as these products complement each other and increase customer utility in concert~\cite{SHN21}. (Incentives in horizontal interoperation can be less cooperative.) Competitive friction in vertical interoperation may arise either from a platform maintaining extractive practices (e.g., by taking commission on in-app purchases; see \S\ref{sec:case-studies:vetting}) or from a platform's efforts to protect its own complementary, higher-layer product that is somehow privileged on its platform---this is called \emph{self-preferencing} (e.g., payment apps having unique access to hardware functionality; see \S\ref{sec:case-studies:hybrid}).  The case where there is an in-house, higher-layer, complementary product with exclusive access to platform functionality is a \emph{hybrid} case in our taxonomy: the platform and the third-party product are complements, and the company's higher-layer product is a substitute. This hybrid type of economic relationship where the platform is able to use its position to shut out other products is uniquely potent and the potential competitors are very vulnerable.

\paragraph{Relation to taxonomy}
The economic context of horizontal or vertical interoperation may be useful to inform contextual analysis of SvI arguments. Security engineering arguments often relate to horizontal interoperation, and security vetting arguments are often, though not always, related to vertical interoperation.\footnote{Theoretically, both types of arguments can arise in both horizontal and vertical interoperation scenarios. In practice, many complex real-world scenarios exhibit aspects of both, but the observation we wish to highlight is that there is a stronger correlation between certain categories.} 
\iffullversion
\else
    See Figure~\ref{fig:bipartite}.
\fi

It makes sense that vetting concerns and vertical interoperation tend to be seen together as they both arise when one company has a platform with which others interoperate. The hybrid cases from our taxonomy are typically also situations in which third parties are pursuing interoperability on a company's platform, so they are also more strongly associated with vertical interoperation.

\definecolor{myblue}{RGB}{80,80,160}
\definecolor{mygreen}{RGB}{160,80,80}
\begin{figure}[ht!]\centering
    \vspace{-20pt}
    \begin{tikzpicture}[thick,
      every node/.style={draw,circle},
      fsnode/.style={fill=myblue},
      ssnode/.style={fill=mygreen},
      every fit/.style={ellipse,draw,inner sep=-2pt,text width=2cm},shorten >= 3pt,shorten <= 3pt
    ]
    
    \begin{scope}[start chain=going below,node distance=5mm]
    \node[fsnode,on chain] (cat1) [label=left: Engineering] {};
    \node[fsnode,on chain] (cat2) [label=left: Vetting] {};
    \node[fsnode,on chain] (cat3) [label=left: Hybrid] {};
    \end{scope}
    
    \begin{scope}[xshift=3cm,yshift=-0.5cm,start chain=going below,node distance=5mm]
      \node[ssnode,on chain] (horiz) [label=right: Horizontal] {};
      \node[ssnode,on chain] (vert) [label=right: Vertical] {};
    \end{scope}
    
    \draw[line width=2.5pt] (cat1) -- (horiz);
    \draw[dashed] (cat2) -- (horiz);
    \draw[dashed] (cat3) -- (horiz);
    \draw[dashed] (cat1) -- (vert);
    \draw[line width=2.5pt] (cat2) -- (vert);
    \draw[line width=2.5pt] (cat3) -- (vert);
    \end{tikzpicture}
    \vspace{-10pt}
    \caption{Our taxonomy and horizontal/vertical interoperation (strong/weak correlations in bold/dashed lines respectively)}
    \label{fig:bipartite}
    \vspace{-10pt}
\end{figure}

\paragraph{Scope of our taxonomy}
Our work focuses on security concerns in the context of adding interoperability to systems that previously had less or no support for interoperation. Some secure technologies have been designed to support open interoperation from the ground up, and some of the types of concerns discussed in this section will not arise at all in such cases. Interoperation between email providers (with PGP) is a good (horizontal) example; third-party software on Linux is a good (vertical) example illustrating that interoperation and security can be built from the ground up.\footnote{Of course, PGP has well documented usability and other issues~\cite{PGP, clark_email}. Our point here is not to extol these examples' security, but to contrast building in interoperation from the outset vs. retrofitting. See also Note \ref{note:hard} discussing how designing secure interoperation is hard.}

\section{Case Studies and Analysis}
\label{sec:case-studies}

Based on recurring themes from the SvI discourse we reviewed, we propose four critical questions comprising an analytical framework for SvI concerns.
    \begin{enumerate}
        \item \qone
        \item \qtwo
        \item \qthree
        \item \qfour
    \end{enumerate}
    
Next, for each taxonomy category---engineering (\S\ref{sec:case-studies:engineering}), vetting (\S\ref{sec:case-studies:vetting}), and hybrid (\S\ref{sec:case-studies:hybrid})---we provide 
\iffullversion
\else
    1--3
\fi
selected case studies,
and analyze each case study with the four preceding questions in mind. These four questions supplement the taxonomy but are not explicitly linked to the structure of the taxonomy; rather, they expand on the core phenomenon and question in each of the categories. Then in Section~\ref{sec:big-picture}, we transition from individual case analysis to comparative analysis, structured around the same four questions.

\subsection{Security Engineering Concerns}
\label{sec:case-studies:engineering}
In security engineering cases,
we have parties trying to build or retrofit their systems to work together. Picture dual-gauge trains with different wheel gauges being engineered to be able to operate (safely) on tracks of different width.  In our cases, the companies involved are generally competitors and the interoperation is generally horizontal.

\refstepcounter{casestudy}
\paragraph{Case Study \thecasestudy: E2EE Messaging}
\label{sec:case-studies:messaging}
This case study discusses two examples of E2EE messaging, iMessage and WhatsApp, which involve distinct SvI considerations.
Here, interoperation means that users of one service could send an end-to-end encrypted (E2EE) message to the users of a different E2EE service---while preserving security.\footnote{Regulation 2022/1925 of the European Parliament and of the Council of 14 September 2022 on contestable and fair markets in the digital sector and amending Directives 2019/1937 and 2020/1828 (Digital Markets Act), OJ [2022] L 265/1, Article 7(3).} 
We encountered mandated interoperability of Messenger (Meta) directly in our case search. There are also similar mandates for WhatsApp in the EU and iMessage in the US.\footnote{Case DMA.100024 - Meta, number-independent interpersonal communications services, (2023)  OJ C, C/2023/1092, 05.09.2023.; \emph{United States v. Apple, Inc.}, No. 2:24-cv-04055, 2025 WL 1829127 (D.N.J. June 30, 2025).\label{fn:cases}}  These cases have many similarities so we discuss them together. We found SvI arguments stating that interoperation is technically difficult and that third parties might not have adequate security guarantees \cite{meta_interop, imessage_svi}.

Before the actions of the competition authorities, WhatsApp and Messenger users could only message other users on WhatsApp and Messenger, respectively. WhatsApp has taken limited measures to support interoperation; Messenger remains siloed. iMessage is only available on iPhones, but iMessage users could always send and receive (SMS) messages to and from Android phones.\footnote{iMessage-to-iMessage communication offers 
\iffullversion
\else
    E2EE
\fi
and additional features not available in iPhone-to-Android communication on iMessage. Unlike iMessage, WhatsApp and Messenger do not support SMS messaging. The three apps are used for similar purposes in practice.}

There is substantial engineering work to be done to achieve secure interoperation across E2EE messaging applications---even for basic messaging functionality, putting aside common auxiliary features such as group messages, disappearing messages, read receipts, and so on. Meta's encryption for both WhatsApp and Messenger is based on the Signal Protocol, which is developed by the non-profit Signal Foundation, and is considered the gold standard in the field \cite{meta_interop}. Meta has stated a requirement that third parties seeking to interoperate must use the Signal Protocol or ``a compatible protocol if they are able to demonstrate it offers the same security guarantees'' \cite{meta_interop}. For smaller providers, the possibility of interoperating with WhatsApp may provide ample incentive to do the engineering work to meet these requirements. For Meta, secure interoperation will also require some engineering, including building secure connections with third-party servers and coordinating
message formats  \cite{meta_interop}. 

In contrast, iMessage does not use the Signal protocol \cite{imessage_v_signal}, and has sufficient market dominance that it may not be incentivized to either switch to the Signal protocol or otherwise facilitate interoperation. Apple has been moving towards adoption of RCS for messages between iMessage and Android users, possibly due to regulatory pressures, and while RCS has many additional features as compared to SMS/MMS, it is not E2EE \cite{apple_bubbles}. Historically, many communication protocols have been agreed upon by broad communities of users and operators, as with email. MLS, based on the Signal protocol, is an effort in that direction and we might eventually have a system of community standards for secure messaging.  
For now, the efforts for greater interoperation consist of some companies retrofitting interoperation conditional on adherence to their standards, and some community-led protocol development efforts. 
There are serious technical and research challenges to interoperation, especially when it comes to the myriad of features available in modern messaging apps~\cite{dma_how_to_e2ee,dma_concerns_e2ee_ross}.

For the incumbents, interoperation entails allowing their services to send and receive information outside of their own networks, and thereby potentially helping competitors. Major incumbent services are currently sufficiently popular that not allowing greater interoperation has been a boon---encouraging greater uptake---rather than a deterrent for customers (e.g.,~\cite{GARCIASWARTZ2019110}).
With the example of Apple and iMessage in particular, the incompatibility of messaging may play a role in users switching to iPhones because iMessage is only available on iPhones \cite{imessage_doj}. 


\subsubsection{Analysis: Security Engineering Concerns}

~\smallskip

\question{1. \qone}
The incumbent companies currently appear to enjoy benefits from locking in users to their platforms (`network effects'~\cite{JULLIEN2021485}), which likely outweigh incentives to interoperate.\footnote{This appears likely from the companies' resistance to recent interoperability mandates; see also, e.g.,~\cite{GARCIASWARTZ2019110} for an empirical perspective.}
Apple can also bundle goods in a way that incentivizes exclusivity. The fact that iMessage is only available on iPhones and is not fully interoperable with messaging from Android phones~\cite{apple_bubbles} may further incentivize users to switch to iPhones.  These incentives may lead companies to avoid full interoperation, rather than to enact exclusionary measures on competitors. 

\question{2. \qtwo} In many cases, interoperation may not be  economically advantageous for the companies being asked to interoperate, as discussed above. For WhatsApp, opening up a closed system is not only an engineering challenge; it also means losing existing network effects. 
For iMessage, extensions to support full interoperation, including encryption, will be an engineering task and will remove an incentive for Android users to switch to iPhones.  Here, however, there are also security arguments in the other direction: encrypting iMessage communications to/from Android better protects iMessage users. This is an interesting example of a company apparently eschewing security in favor of other economic incentives, yet at the same time framing some of their arguments for doing so as security arguments.

\question{3. \qthree}
Although Meta's messaging apps already build on the Signal Protocol for WhatsApp, there is still a substantial amount of engineering to make sure that interoperation is secure. The Signal protocol is a centralized server design, the predominant paradigm in encrypted messaging, and Meta has also made some modifications to the Signal protocol in WhatsApp \cite{meta_interop}. For Apple, some of the changes that are necessary for interoperation are simpler (such as chat bubble color~\cite{apple_bubbles}) whereas some are more complicated (such as encryption). In both cases, using community standards---such as Messaging Layer Security (MLS)---might achieve strong security for users and easier interoperation in the future.

\question{4. \qfour} 
End-to-end encryption is a well-researched goal, and there are standards bodies and academic researchers as well as companies looking to improve the existing technology. There are serious open questions around how best to achieve interoperation in E2EE, especially between existing systems that need to be `retrofitted' with interoperation capabilities; at the same time, there are promising avenues using existing technology to set up interoperation that remains secure~\cite{dma_how_to_e2ee,dma_concerns_e2ee_ross}. Thus, it is an open question in which contexts security considerations may be an insurmountable obstacle to an interoperability mandate for E2EE messaging. Research progress over the coming years may reveal new challenges and/or ways to do interoperable E2EE better.

\subsection{Security Vetting Concerns}
\label{sec:case-studies:vetting}
Next, we present three similar relevant technologies, featuring distinct SvI concerns 
in the vetting category. All three case studies relate to Apple app stores as this has been the focus of EU efforts. There have been similar US cases against both Apple and Google (see Note~\ref{note:apple}).

We begin with some context. Apple has historically adhered to a `walled garden' approach for iPhones. Users are only able to download and use apps vetted by Apple according to their security, privacy, and quality standards. The stringent screening of apps on the App Store has furthered the iPhone's reputation for strong security\cite{blog_android_apple_security}. 

This approach limits what users can do on their own mobile devices---far more strictly than on any existing PC systems.\footnote{Recall that we use ``PC'' broadly to include Macs; see footnote~\ref{fn:pc}.}  Apple's own approach on its Macs contrasts with its approach towards third-party software on its other devices. Apple has a notarization process for software intended for Macs, but this process is automated and much more limited than the process for iOS apps \cite{apple_dev_mac, apple_app_review}. It is also cheaper---the cost to developers for Macs is membership 
in the Apple Developer Program (99 USD yearly) \cite{apple_dev_license_agree}, which iOS developers must have in addition to other substantial fees.\footnote{See the Appendix for more detail on fees.} Once notarized, Mac software can be distributed via a website, not only via Apple marketplaces. These policies reflect Apple's discretionary choices about tradeoffs between user choice and security vetting on different products. 

From the perspective of developers and antitrust officials, there is another dimension to this tradeoff.  Apple's vetting of apps and its position as gatekeeper and security guarantor coincides with a position of market power.  Apple's control over its App Store allows it to determine which apps are offered and to take a cut of transactions. 

Thus, to cast security vetting concerns as only a balancing of user choice against security, is to miss the economic incentives at play.  The core reasons behind the push towards more open app distribution appear to have been economic. Developers are not arguing for the freedom to distribute malware; rather, they want to avoid the substantial rents that gatekeepers force upon them.\footnote{Findings of Fact and Conclusions of Law Proposed by Epic Games, Inc., \emph{Epic Games, Inc. v. Apple, Inc.}, 559 F. Supp. 3d 898 (N.D. Cal. 2021), \emph{aff'd in part, rev'd in part}, \emph{Epic Games, Inc. v. Apple, Inc.}, 67 F.4th 946 (9th Cir. Apr. 24, 2023).}

\refstepcounter{casestudy}
\paragraph{Case Study \thecasestudy: App Sideloading}
App sideloading refers to downloading an app onto a device via some route other than the official app store. This alternate route might be a website or alternative app store. Sideloading was not allowed on iPhones (or iPads) until the DMA, and is still not permitted in most places outside the EU. Apple still requires that sideloaded apps be notarized, meaning the app is vetted and signed by Apple and only apps with such a signature can be downloaded onto a device.  This notarization procedure is not as involved as the vetting for the App Store, but still includes all the security requirements in the App Store Guidelines \cite{apple_app_review}. A major reason why Apple says they need to notarize all apps, especially those with alternative distribution, is to maintain the security standards of apps on its App Store \cite{apple_white_paper}.

The notarization process is not necessarily cheaper than getting approved for the App Store, despite there being fewer requirements for Apple to check.  Developers that choose sideloading must pay a 0.50 EUR `core technology fee' (CTF) for each first annual install over one million, where annual installs include updates \cite{apple_web_sideloading}. Apple allows free installs for a year by the same user, so within a year, the app can be updated multiple times for free, but there is a charge to update each following year \cite{apple_ctf}. Updating apps to fix security issues is an important means of protecting users.  As timely updates are important for security, disincentivizing proactive updates may undermine Apple's stated goals of preserving security.  Depending on a developer's circumstances, this alternative pricing is not necessarily cheaper than the 30\% cut Apple takes from the App Store and in-app purchases.\footnote{See Appendix for an example calculation of fees.}
Apple justifies these fees by invoking the role it plays in developing tools for developers, which include security benefits from the notarization process \cite{apple_ctf, apple_press_release_business}.

In addition to paying fees on purchases of apps, Apple requires all \emph{in-app} purchases (IAP) of digital goods to be intermediated by Apple, and generally subjects these to a 30\% fee  \cite{apple_membership} (this fee structure has been changing in some places as a response to competition enforcement, including in the EU).\footnote{Even developers who choose a link out option now available in the EU still have to pay a 10-17\% commission on sales of digital goods and provide Apple with transaction reports \cite{apple_alt_payment_eu}.}  Before the DMA and other efforts, Apple also forbade apps from linking to websites where purchases could be completed like any other online purchase (without Apple intermediating) \cite{apple_app_review}. Different rules govern the purchase of physical goods. The type of good (physical or digital) being purchased would appear to have no bearing on payment security. However, Apple's justification for requiring usage of Apple's IAP system is to ensure the security of transactions\cite{apple_white_paper}. That said, intermediaries for processing payments securely online are common, and usually levy much smaller fees (e.g., Epic reports less than 5\%).\footnote{\emph{Epic v. Apple}, \emph{supra} note 28, at \textparagraph 454.} 

\refstepcounter{casestudy}
\paragraph{Case Study \thecasestudy: Alternative App Stores}
\label{cs:app-stores}
Apple must now allow alternative app stores, which will be able to distribute sideloaded apps. As discussed above, the apps being sideloaded must be notarized by Apple. 

The alternative app store itself must abide by additional criteria set by Apple, which Apple can check before allowing the alternative marketplace to operate. Alternative marketplaces must, for example, agree not to distribute apps that infringe intellectual property rights or that are malicious as well as deal with requests for removal of apps by governments\cite{apple_alt_market}.  
Apple also points out that collecting data about apps through its App Store is an important aspect of keeping users secure \cite{apple_white_paper}. There is no reason that interoperability regulations would prevent Apple from engaging in or requiring this type of data sharing with alternative app stores \cite{apple_white_paper}. Indeed, this type of data sharing would be beneficial even across existing marketplaces outside of Apple's control, e.g., Google Play. 

The vetting concerns have been the prominent ones in these cases, but there are also engineering considerations involved.
Apple created an API to allow these alternative marketplaces to be able to execute downloads \cite{apple_alt_market}.  There is also some extra work that needs to be done for developers distributing their apps via their website\cite{apple_app_distribution_website}.

\refstepcounter{casestudy}
\paragraph{Case Study \thecasestudy: WebKit}
Prior to the DMA and the subsequent actions taken by the Commission, all web browser apps on iOS had to use WebKit as their browser engine \cite{apple_webkit_requirement}.  Browser engines are part of web browser apps (and in-app browsing) that render the content of a website into a page ready for user interaction \cite{browser_sok}.  WebKit is open source, but belongs to Apple\cite{apple_webkit_owns}. As a result of Commission actions, Apple now allows alternative browser engines in the EU, both for browser apps and in-app browsing \cite{apple_alt_webkit}.\footnote{Apple more recently allowed alternative browser engines in Japan after recent competition legislation~\cite{apple_jp_webkit}. 
See Section \ref{sec:generalizability}.}  
Prior to these changes, Apple’s iOS was the only operating system that did not support other browser engines; other browser engines can be used on the operating systems of their competitors~\cite{team_gb_browser_engines}. In the developer documentation, the requirement that browsers use WebKit is not explicitly based on security, although there is notable community discussion around security motivations for this requirement~\cite{yes_we_cite_reddit}. 

Browser engines play a key role in preserving security for users online. Here, iOS users may be disadvantaged by the scarcity of options available \cite{browser_security}.  WebKit has not been demonstrated to be, and does not have a strong reputation for being, any more secure than other browser engines \cite{browser_security}. There have been bugs in WebKit and Safari \cite{browser_sok}, and not allowing developers (of browsers or those using in-app browsing) to use the tools they see fit could constrain their ability to make application-specific security choices.

Apple stipulates that any developer looking to use an alternative browser engine must meet certain requirements, including some related to security \cite{apple_alt_webkit}.  Any browser app would also have to abide by Apple's general app guidelines, including those regarding security \cite{apple_app_review}.  
Any developers looking to implement an alternative browser engine must be approved by Apple, allowing Apple to vet potential alternatives and protect users from insufficiently secure options \cite{apple_alt_webkit}.

\subsubsection{Analysis: Security Vetting Concerns}

~\smallskip

\question{1. \qone}
The three case studies share the important context of Apple's control over third-party developers' access to their potential customer base on the iPhone. iPhone and iOS comprise a relatively closed device and platform ecosystem, and Apple can exclude developers on grounds that their apps do not meet Apple's guidelines or because their services are not allowed in the ecosystem (e.g., alternative app stores). Apple's monopolistic position definitionally also allows it to charge developers more than would be expected in more competitive settings. This type of `gatekeeper' power differs markedly from the market power discussed in the security engineering case study above, and in cases involving horizontal interoperation more generally. Here, Apple has the power to completely exclude a third party; in horizontal interoperation, it cannot cleave third-party developers from users.  

\question{2. \qtwo} 
Vetting apps to protect user security is a reasonable end, but it is also used as a means to extract rents. Keeping malware and other insecure apps away from users is also good for platforms, especially those that, like Apple, are reputed for security and curation \cite{blog_android_apple_security}.  It is hard to (externally) evaluate whether the price tag is justified, given system opacity and limited competition. 

\question{3. \qthree}
In all these cases, some amount of vetting is important to preserve the security status quo. Apple's notarization system and CTF is Apple's suggested way forward. As discussed above, the notarization guidelines are less onerous for Apple to check than their status quo, and vetting new app stores will likely be a much rarer occurrence than vetting new apps---but Apple's new mechanisms are not necessarily cheaper for developers. There have been changes to Apple's pricing practices (and it is possible more changes will need to be made in the future).  Price changes might be impactful for the company's profits, but changing pricing is typically not an engineering or significant vetting challenge.

\question{4. \qfour} 
Apple has absolute control over third-party developer access to iPhone users. At the same time, they rely on those third parties to develop apps for the iPhone, increasing the iPhone's utility and commercial success.
The ecosystem of iOS apps is massive, and preventing every possible `bad' outcome is too general and too far-reaching to be a viable reality.  Instead, Apple has policies (e.g., \cite{apple_app_review}) around how they protect users and devices `well enough.'  Inherent in these policies are tradeoffs around security, safety, privacy, and quality.  Too-stringent guidelines might unnecessarily limit available apps or take too many resources.  Too-lenient guidelines might undermine user trust and device integrity. 

The changes the Commission is enforcing will not necessarily impose new tradeoffs when it comes to app vetting.  Apple itself decided to apply only a subset of app guidelines to those apps applying for notarization (whether for economic reasons or otherwise).  Alternative app payments and alternative app stores will require users to invest their trust in other third parties.  Then, choosing an acceptable tradeoff falls in larger part to the users; they must choose which companies' security (and privacy) policies are trustworthy and/or worth the risk.

\subsection{Hybrid Concerns}
\label{sec:case-studies:hybrid}
Recall that hybrid cases are generally those in which the company is opening up a functionality to third parties that was previously only available to its own in-house product.  To do this, it must engineer a system that can preserve security when opened up to third parties, as well as vet those third parties.  

\refstepcounter{casestudy}
\paragraph{Case Study \thecasestudy: NFC for mobile payments}
We identified Near Field Communication (NFC) as a relevant technology from both the Apple Mobile Payments antitrust case\footnote{Case COMP. AT.40452 Apple - Mobile Payments, (2024) OJ C, C/2024/1027, 19.01.2024.} and the Qualcomm/NXP Semiconductors merger.\footnote{Case COMP. M.8306-Qualcomm / NXP Semiconductors, (2018) OJ C 113, 27.3.2018, p. 8.} The Qualcomm/NXP Semiconductors merger mandated interoperation between NFC chips and other chips whereas the Apple case mandates interoperation between those types of chipsets and software. (There is also a similar NFC-related antitrust proceeding in the US against Apple.)\footnote{\emph{United States v. Apple}, \emph{supra} note 23.} In our search for SvI concerns, we did not find any articulated concerns regarding the merger so we focus on the Apple case. The `security vs. interoperability' considerations discussed by Apple include that new APIs had to be built to allow third-party developers secure access to the technology \emph{and} that those developers will be vetted before such functionality is available \cite{apple_se_press_release}, which falls into our hybrid definition. 

NFC is a set of protocols that work at close range and allow two devices to communicate securely.  NFC is used for myriad applications linking smartphones to other devices, including reading tags on objects, unlocking doors, or communicating credit card or identification data \cite{apple_security}. For an app to use NFC for mobile payments, it must have access to the phone's NFC antenna. It also needs to access key material stored on the Secure Element, which holds sensitive information, including the card information necessary for host card emulation (HCE) \cite{apple_se}.  Apple had restricted third parties from using NFC and the Secure Element for HCE, reserving this ability for Apple Wallet and Apple Pay. It did allow other usage of NFC via API access to the necessary hardware and software \cite{apple_hce}. 

EU competition authorities identified this as monopolistic behavior and Apple has opened up the HCE capability to third-party apps, subject to approval by Apple. The approval criteria require third-party developers to commit to complying with laws and standards related to privacy and security, including standards set by EMVCo, a standards body for card-based payments \cite{apple_hce}.  In addition, Apple requires third-party developers to have certain policies around privacy and dealing with vulnerabilities~\cite{apple_hce}. This is a vetting concern, and screening developers seeking to use HCE is an important aspect of protecting security in the newly open system.

In order to `open up' NFC capabilities for third-party wallet applications, third-party developers must have access to NFC, which is already available, as well as access to the Secure Element \cite{apple_se_press_release}.\footnote{Android supports HCE with or without a secure element, and allows third-party interoperation with the latter method; third-party Android developers cannot access the secure element \cite{android_hce}. If Apple adopted a similar approach, it could potentially achieve interoperability without involving the secure element at all; it has not opted for this approach.}  Access to the Secure Element is already separate from from the Secure Enclave, which is where the operating system stores other sensitive data \cite{apple_security}.  This new access must be built correctly in order to preserve the security of both the third-party data and the rest of the system.
The Secure Element has internal security boundaries that could impede a malicious developer from accessing other sensitive data and none of this data needs to come into contact with the operating system or app---the reply can go straight from the hardware to the payment processor \cite{evil}.  Thus, APIs built for third-party developers could grant access to only specific functionalities without compromising others.  There are also existing vulnerabilities in these products \cite{evil}, which would not necessarily be exacerbated by interoperability and might benefit from greater attention from the community.  

\refstepcounter{casestudy}
\paragraph{Case Study \thecasestudy: OS and Physical devices}
We identified OS and physical device interoperation as a type of relevant technology from three cases: two similar cases involving Apple's iOS\footnote{Case DMA.100204 SP - Apple - Article 6(7) process, (2025) OJ C, C/2025/5000, 18.7.2025 and Case DMA.100203 Apple – Operating systems – iOS, Article6(7), SP, Features for Connected Physical Devices, (2025) OJ C, C/2025/4646, 14.8.2025.} and the Google/Fitbit merger.\footnote{Case COMP. M.9660 Google/Fitbit, (2021) OJ C 194, 21.5.2021, pp. 7–16.} The Apple cases concern the interoperability between many aspects of iOS and third-party hardware and software whereas the merger focuses on the interoperation between specific third-party hardware and the Android OS. (The US case concerning NFC, mentioned above,\footnote{\emph{United States v. Apple}, \emph{supra} note 23.} also discusses physical device interoperation in a similar light to the EU cases about Apple.)  The SvI arguments regarding the Apple cases are exemplified in a whitepaper by Apple~\cite{apple_white_paper_2}.  

Public facing discussions about security in the Google/Fitbit merger case were notably very different; they did not discuss the security of the \emph{mandated} interoperability, but rather discussed distinct \emph{voluntary} interoperability that would result from the merger, reassuring users that it would not endanger user security or privacy \cite{fitbit}. This type of argument is outside our current scope, so we focus on the Apple case for the rest of the case study. However, investigating this type of `inverted' case---and examining how perspectives on the the compatibility of security and interoperability feature differently across economic contexts---could make for interesting future work.

The Apple cases cover many different hardware and software features of iOS, including notifications on smartwatches, background execution, content casting and transfers, and paired device set up.  Apple will be required to provide interoperability with the full list of functionalities to third parties.  All of these functionalities are already available to Apple's own devices (including Apple Watch, AirPods, and other iPhones), so achieving interoperability is a question of opening up previously reserved functionality to third parties, generally via APIs and SDKs.

Although the stipulations from the Commission are still evolving, Apple is currently evaluating interoperability requests on a case-by-case basis \cite{apple_white_paper_2}.  In their recent public-facing document, Apple mentions that Meta has made 15 requests for interoperation regarding various functionalities \cite{apple_white_paper_2}. While Apple may be required to interoperate with competitors, the regulators do not require that any interoperation request be granted in a way that would undermine the security of the device.\footnote{Regulation 2022/1925 of the European Parliament and of the Council of 14 September 2022 on contestable and fair markets in the digital sector and amending Directives 2019/1937 and 2020/1828 (Digital Markets Act), OJ [2022] L 265/1, Article 6(7).} Thus, Apple could deny interoperation requests from known malicious actors and develop/require bespoke solutions for interoperation to reduce security risk.

One example is peer-to-peer (P2P) Wi-Fi connections, which can be used to transfer files, such as with Apple's AirDrop, and continuity services, such as Handoff or Universal Clipboard\cite{Darmstadt_21}. These functionalities use a Bluetooth Low Energy (BLE) connection between two devices to create an \emph{ad hoc} P2P Wi-Fi connection, which uses Apple's iCloud credentials as part of its certificate procedure \cite{apple_security}. Several aspects of these services, including BLE, have been shown to leak user information or otherwise have vulnerabilities \cite{BLE_not_private, ble_not_private_2, apple_bluetooth}.  Apple Wireless Direct Link (AWDL) is an key protocol used in P2P Wi-Fi, which is based on IEEE 802.11, a set of standards, but is currently proprietary \cite{Darmstadt_18}.  This protocol has been shown to have security vulnerabilities \cite{Darmstadt_18, Darmstadt_19, Darmstadt_21}. AWDL being proprietary is a barrier both to better study of its security and to interoperation \cite{Darmstadt_18}. 

To open P2P Wi-Fi up more broadly, the Commission proposes\footnote{Case DMA.100203 Apple – Operating systems – iOS, Article6(7), SP, Features for Connected Physical Devices, (2025) OJ C, C/2025/4646, 14.8.2025.} that Apple either make AWDL available to third parties or facilitate third-party interoperation through Wi-Fi Aware,\footnote{Also called Neighbor Awareness Networking (NAN).} which is a public protocol with similar functionality that is based on AWDL\cite{wifi_patent}. If Apple makes AWDL public, there is little engineering to be done, if the protocol is indeed secure.  If the security of AWDL is reliant upon its details remaining secret, i.e. \emph{security through obscurity}, Apple and its users might be better served by Wi-Fi Aware or an improved version of AWDL. 

\subsubsection{Analysis: Hybrid Concerns}
The hybrid cases are more than just the sum of the other two; we see (arguably more complex) patterns specific to this category.  

\question{1. \qone}
The platform's market power is again a key feature of these cases. Since third-party products compete directly with those of the incumbent company (e.g., Apple Watch competing with third-party watches), the company has incentive to protect their current market share.
At the same time, the company has control over access to those key functionalities and can block access for competitors.
In contrast to vetting concerns, interoperation means that third parties gain access to functionality that was not previously available to third-party developers at all.  
Since the company must both build access to previously exclusive functionality and also vet who gets that access, the platform's control over the situation is broader than in either engineering or vetting cases.  Thus, the company's ability to block access may be more robust in the hybrid case.

\question{2. \qtwo} The platform has a clear economic incentive to protect their exclusive products that are bundled with or complementary to their main devices. Preventing third-party developers from offering the full functionality available on Apple's own products reduces the  value of those other products and puts theirs at an advantage, i.e., self-preferencing. Supporting interoperation would require security engineering in concert with third parties.  These incentives may coalesce into significant motivation for a platform to keep systems closed. 

\question{3. \qthree} 
 Not only must the platform build secure APIs or other means of access for third-party developers, but many of these APIs may need to be bespoke for specific uses. This could be a significant engineering task. The platform can vet all proposals for interoperation; this enables them to enforce security standards, and  keep tabs on what third-party developers are doing. Given the previously exclusive nature of the functionality being opened up, interoperation in hybrid cases may be subject to more scrutiny from the platform in control than in vetting cases, and may require more coordination between the platform and third parties. Although perhaps costly, this may make for better, more secure products.

\question{4. \qfour} 
Existing interoperation policy involves interoperation with third parties, despite some potential risks (such as with Bluetooth devices \cite{bluetooth_mitm}).
When opening interoperation, the platform's greater case-by-case control in hybrid cases may mean less difficulty in maintaining security standards than when having to vet millions of apps as in the pure vetting case.

\section{Comparative Analysis and Takeaways}
\label{sec:big-picture}
We now offer a \emph{comparative} analysis, identifying key considerations across our taxonomy and case studies. First, we consider \emph{economic incentives, market power, and security}, corresponding to Analysis Questions 1 and 2. Then we discuss \emph{security trade-offs}, corresponding to Analysis Questions 3 and 4. Finally, we offer big-picture takeaways. Table~\ref{tab:big_table} (Appendix) provides a succinct summary of our comparative analysis.

\paragraph{Economic incentives, market power, and security}
Recognizing the nature of a company's market power can be a key step in understanding the ways in which security can become subordinated to other business goals.
There are two main types of market power that are relevant in the cases we discuss: traditional market dominance, where a company is simply so large it dominates some market; and platform power,\footnote{The precise definition of \emph{platform power} is the subject of some debate in legal and economic literature (e.g., \cite{evans2011platform, cohen2017platforms}). We use \emph{platform power} to mean the power of platforms over third-party products or services that may run on their platform. This is similar to the notion of a company acting as a \emph{gateway} as defined in DMA Article 3(1b).} where incumbent companies are `gatekeepers’ of third parties' access to customers.  We did not specifically categorize cases by the nature of the incumbent company's market power, and yet different security concerns seem to correlate with different types of market power.
Cases with \bi{security engineering concerns} are also those in which the market power of the incumbent is simply \emph{market dominance}.  The relative size of a company, or, more precisely, its market share, and the resultant network effects, can confer significant power without more complicated `gatekeeper' dynamics. As noted in Section \ref{sec:framework}, these instances are also often those in which the necessary interoperation is horizontal, meaning third parties might be on somewhat more equal footing when it comes to making technical decisions.  Interoperating securely is generally an engineering task to be coordinated between the companies involved and, if there are existing standards, it may be clearer what it takes to `interoperate securely.' 

We can see \emph{platform power} clearly in \bi{security vetting concerns}, which are also instances of vertical interoperation.  As noted in Section \ref{sec:case-studies}, vetting concerns involve company discretion regarding security policy choices, and rarely have clear-cut engineering answers. As such, we view narratives around vetting concerns as potentially more easily tailored to fit a company's economic needs. The platform position in these cases typically gives the company significant power to impose potentially punitive prices or exclude third-party developers.  Thus, this type of concern should be identified as particularly susceptible to a blurring of boundaries between security and economic incentives, compared to  engineering concerns.  
The \bi{hybrid case} is not simply a dilution of the dynamics of the other two cases.  \emph{Platform power} manifests here too.  Rather than using platform power to extract fees, as with security vetting concerns, hybrid cases more often feature or involve an incentive for companies to use their platform power to self-preference their products, hobbling or entirely excluding potential competition.  In both vetting and hybrid cases, the company acts as a gatekeeper. The extensive control of incumbent companies to be able to block competition through both vetting and engineering is distinct to hybrid cases, meaning platform power is particularly entrenched. 

The engineering demands in a hybrid case appear in general to be no less significant than in a security engineering case. Thus, platform power is bolstered by legitimate engineering concerns. This means there are two lines of defense to keep out both bad actors and competitors, compounding the effects we see in pure engineering or vetting cases.   Third-party competitors may be more completely shut out in hybrid cases as compared to the other two categories. Without interoperation, messaging apps can still function with their own users, and apps can be distributed on iPhones as long as they adhere to Apple's guidelines and fee schemes. But without access to the HCE, there can be no alternative to Apple Pay on the iPhone. Third-party smartwatches can function without access to all hardware functionalities, but their utility may be significantly limited. Thus, hybrid security concerns should be evaluated with careful consideration of these extra potent market ramifications as compared to engineering or vetting concerns.

\paragraph{Security tradeoffs}
Security engineering involves many practical tradeoffs: between security and efficiency, usability, profit, and more.  Achieving usability and security at once is well known to be challenging~\cite{usability_book}. Usability also interacts with companies' economic incentives: a phone with such tight security that it is unusable would be hard to sell.  
Economic interests can also motivate pressures against interoperation. Just as undermining usability through a narrow focus on security can ruin a product, undermining competition through a narrow focus on security can ruin a market. 
The simplest case is again the \bi{security engineering concern}. Inherent in the idea of an engineering concern is the idea that the problem can be addressed by engineering. The tradeoff, then, relates to how much to invest in trying to solve the engineering problem and how much of a burden that is. In practice, of course, even the best, well-funded systems have bugs and vulnerabilities. Thus, companies must deploy systems that might be vulnerable.  This tension carries over into interoperation, where, as in most other contexts, systems might have vulnerabilities despite their best efforts.

When companies raise \bi{security vetting concerns}, they may argue that security considerations necessitate limitations on interoperation, or perhaps preclude interoperation entirely. Recall that vetting is about \emph{interoperability policies}, rather than \emph{technical interoperability}. Policies generally involve discretionary tradeoffs between security and other interests. The tradeoffs made in existing policies are an essential reference to evaluate how much of a divergence from existing security practices, and how much of an economic burden, a proposed policy change would be.  
The \bi{hybrid case} may feature all of the preceding types of tradeoffs. Yet the hybrid tradeoffs are of a slightly different character than the engineering and vetting trade-offs. Those seeking to interoperate with OS functionality reserved for device manufacturer use are fewer than those seeking to make apps for smartphones, so vetting could be a project of a substantially smaller scale. Relatedly, interoperation could be bespoke, involve more monitoring, and require less scalability. 

\paragraph{Big picture}
Our taxonomy, case studies, and analysis highlight how the following prominent themes feature in `security vs. interoperability' discourse in impactful competition policy contexts in the wild.
\iffullversion
\else
    First, interoperation and economic incentives often conflict.
    Second, security arguments and economic incentives often align.
    Finally, such security arguments and their implications can only be fully understood in the broader economic and policy context as well as the technical security context.
\fi

Where vetting is the primary security concern, incumbent companies necessarily make tradeoffs and also have strong platform control that can be leveraged to extract fees. The hybrid combination of platform power and guarded functionality can grant incumbent companies even more power to exclude direct competition.

\smallskip \noindent \textbf{So, when is interoperation worth it?}
Our framework aims to provide analytical tools to make policy choices around which kinds of interoperability should or should not be pursued, rather than prescribing where that line should be. While our framework addresses some of the economic context, decisions about remedies are made by regulatory bodies after extensive market research and discussion with stakeholders. 
Broadly, the goal should be to mandate interoperation only when the benefits to the market ``outweigh'' the potential security difficulties and risks, which is a challenging question because it requires weighing benefits and risks of incomparable types.  While security experts can help other decision makers better understand this tradeoff, which is the goal of this paper, this is ultimately a policy question that depends on economic, legal, and technological (security) considerations.  Our hope is that judges, policymakers and policy experts will take our framework into account when weighing interoperability and security concerns in technology settings.

\section{Generalizability}\label{sec:generalizability}

Our methodology focuses on the EU, and our discussion makes note of some relevant US cases. Some cases in other jurisdictions bear significant similarity to those in the EU, often addressing the same technologies from the same companies, while others are markedly different. Many countries have completely different companies, market structures, legal systems, transparency of proceedings, and policy development processes. Nonetheless, we are seeing antitrust-based efforts to encourage or mandate interoperation across many diverse jurisdictions, and including several examples where SvI arguments have been raised. A natural question then arises: to what degree can our framework be applied beyond our present context?

While our systematic analysis is limited to the EU, we believe our taxonomy and analytical framework as well as many of our high-level conclusions generalize, offering a helpful starting point for the study of SvI tensions across jurisdictions. This initial groundwork will likely need to be supplemented by jurisdiction-specific considerations --- a ripe area for future work. The foundations that many of our findings are based on are jurisdiction-agnostic, such as the structure of the technology to be interoperated, different stakeholders' incentives to interoperate, and security analysis. 

Based on informally following relevant antitrust proceedings in non-EU jurisdictions (most notably the US, but also several others), we have observed numerous cases from other jurisdictions that fit neatly into our taxonomy and exhibit some of the same incentive patterns seen in our EU examples --- while, at the same time, featuring market structures, power dynamics, and technologies that differ interestingly from our EU cases.
We briefly mention a number of noteworthy examples (from China, the UK, Brazil, and others) below. We stress that we have not thoroughly examined these cases and that we have no expertise in legal systems beyond the EU and US.

In 2021, Chinese regulatory authorities started to push large superapps\footnote{Superapps are applications with many services in one app (such as messaging, payments, ride hailing, and social media).  Many have third-party miniapps that are apps published inside the super-app \cite{superapp_sok}.} WeChat and Alipay to facilitate greater interoperation between the many services that these two superapps provide, for example asking WeChat to make Alibaba-owned e-commerce links more usable in its messaging app.  The companies have suggested that security is one of the reasons their app did not previously support such links \cite{scmp2021walledgardens}.  This appears to be a vetting case that is not about apps on operating systems (unlike our EU examples), but exhibits many similar characteristics, including the fact that SvI arguments have been made by platforms that mediate other companies' access to users.  What is different about this case is the complicated and interconnected economic relationship between Alipay and WeChat, where each acts as an upstream platform to the other in different markets~\cite{superapp_sok}, a phenomenon distinctive to superapps. 

In the UK, a 2017 mandate required large banks to build standard APIs to allow third parties access to customer data \cite{UK_banking}. Banks raised hybrid SvI concerns \cite{collinson2018open, barclays_open_banking}, but the mandates were ultimately implemented. In another similar effort to spur competition in the financial sector in Brazil, the central bank itself built the transaction infrastructure, called Pix \cite{pix_BCB}. Stakeholders raised hybrid security concerns, discussion of which is ongoing; some have been addressed after the roll-out~\cite{pix_fraud,pix_rules}. Here, it fell to the central bank to support secure interoperation, setting it apart from other cases where this burden falls largely on the private parties expressing SvI concerns.

There are also instances where the incumbent companies are already performing systematic vetting, and the interoperation mandate, in essence, is asking that the results of this vetting be shared with competitors.  We see this in one of the proposed \emph{Epic v. Google} remedies in the US, which suggested requiring Google to share their app catalog with potential third-party app stores.\footnote{Epic Games, Inc. v. Google LLC, No. 24-6274 (9th Cir. July 31, 2025)} Google pushed back with multiple arguments, including the idea that this would result in third-party app stores freeloading on Google's security vetting.\footnote{Google LLC's Proffer Regarding Epic's Proposed Remedies, Epic Games, Inc. v. Google LLC, No. 3:20-cv-05671-JD (N.D. Cal. June 24, 2024), ECF No. 671.} Google also raised concerns about building the necessary infrastructure to allow for catalog access and downloads, making this a hybrid argument but the remedy was not ultimately pursued \cite{settle_epic}.  Korean superapp Naver made a similar freeloading argument regarding its vetting of real estate listings, when it was asked by competition authorities to change contracts that prevented competitors from using the same listings~\cite{lee2020naver}. Here, interoperation is not pitted directly against security, but instead we see a pattern of mandating competitor access to vetting results in order to support secure interoperation, which naturally raises free-riding concerns. 

In addition to the catalog example in the US discussed above, many jurisdictions have taken varying approaches to dealing with the same issues we see in the EU, such as Japan \cite{japan} and India\cite{india}.  

For example, Japan's Act on Promotion of Competition for Specified Smartphone Software includes provisions regarding alternative app stores, in-app payments, link outs, and browser engines that are similar to the relevant EU efforts~\cite{japan_jftc_act}.  A notable difference is that there is no requirement about sideloading from the web, meaning the act requires that alternative app stores be allowed, but not that third-party developers should be allowed to disseminate their apps directly to users from a website~\cite{vandewalle2025japan}. This distinction is in part due to security and safety concerns~\cite{vandewalle2025japan}.

\section{Other Open Directions}

\paragraph{Security, privacy, and safety}
The terms \emph{security}, \emph{privacy}, and \emph{safety} are distinct but often used in overlapping ways \cite{ross_book}. The SvI arguments we document often include elements of privacy or safety; but unlike the other two, security was a consistent core element in all the SvI-style arguments we encountered, and our primary focus. Untangling and analyzing security, privacy, and safety threads within the discourse could be fruitful future work.

\paragraph{Security vs. openness}
Our framework may offer insights regarding ``security vs. openness'' arguments more broadly --- for example, in the contexts of data portability, content moderation, or API access for agentic AI models. 
The idea that \emph{openness often benefits security} has long been established in computer-security community, as well as the idea that ``security by obscurity'' arguments may be favored to conceal poor security practices or protect proprietary interests~\cite{HMP25}.
A full exploration of this connection is beyond our scope, but would be interesting future research.

\section{Conclusion}
\label{sec:conc}

We have presented a systematization of ``security vs. interoperability'' arguments based on an analysis of EU cases. We offer a taxonomy, a analytical framework, and case studies of the diverse contexts in which SvI arguments arise; our comparative analysis underscores their distinguishing characteristics.

Court proceedings and regulatory investigations in the area will continue to be common~\cite{bloomberg-stolton}; many of the investigations we mention are ongoing (e.g.,~\cite{eu_alt_app}).
There is thus a pressing need to build a nuanced interdisciplinary understanding of the interplay between security and interoperation; only with this full picture can informed trade-offs be made. 
Interoperability mandates are designed to influence economic incentives, which in turn influence security-relevant design decisions; security arguments are generally embedded within broader economic, political, regulatory interactions where technical considerations are just one factor among many. 
We hope our contributions provide an actionable framework for deeper interdisciplinary engagement between the security community and lawyers, regulators, and policymakers to clarify and mitigate potential security harms in the context of antitrust.

\ifanon\else
    \section*{Acknowledgments}
    The authors would like to extend special thanks to Mallory Knodel for helpful discussion and many insights. We are also grateful for the generous help of Alexander Heinrich, who provided detailed consultation on certain technical issues. 
    Thanks also to 
    Kyle Hogan,
    Michael A. Specter,
    Daryl Lim,
    Bryan Choi,
    Stephanie Chen,
    Naveen Rajan,
    Sangyun Lee,
    Andrés Fábrega, and 
    Mathy Vanhoef.
\fi

\section*{Open Science Statement}
\paragraph{Data Availability}
We have no code or other research artifacts to which the conference's open science statement might apply. All our sources are public and cited here.

{\bibliographystyle{acm} \bibliography{refs}}

\section*{Appendix}
\subsection*{Fee calculations comparing App Store Fee and CTF}
We briefly illustrate that the cost of paying the core technology fee (CTF) is not necessarily less expensive than the current 30\% commission fee. Suppose there is an app that costs €1 and has no in-app purchases available.  If this app sees 6 million downloads in one year, then the fee paid to Apple would be $6\cdot0.3=1.8$ million euros if it were distributed by the App Store, but would pay $(6-1)\cdot 0.5 =2.5$ million euros if it were distributed separately and paid the CTF (which is free for the first million downloads) \cite{apple_ctf}.  This constitutes a loss for the developer already in the first year.  If they were to then update the app, which would be freely available to the users, after one year, they would have to pay another €2.5 million while reaping no income. 

Apple also has specific requirements of the developers themselves that must be met for Apple to allow direct web distribution.  Web distribution, i.e. enabling users to directly download apps from the developer's website, is a simple way for developers to have direct access to their users. In order for a developer to be granted permission to do web distribution Apple requires that developers be members of Apple Developer Program for at least two years and have an app that had more than one million first annual installs in the previous year \cite{apple_web_sideloading}.  This would prevent new developers from offering their first apps in this manner.  Presumably, developers that have recently had an app with at least one million first annual installs are likely to follow-up with a similarly popular app.  As with all alternative distributions, the new app would incur the CTF for all installs above one million. Apple has some exceptions to their fees for small developers and NGOs \cite{apple_ctf}, but these groups are less likely to fulfill the eligibility requirements. Thus, while these requirements are ostensibly to control for responsible app developers, they track closely with criteria for selecting only those developers who will also have to pay substantial fees. 

\begin{table*}[htb]
\centering
\renewcommand{\arraystretch}{1.5} 
\begin{tabularx}{\textwidth}{|X|X|X|X|}

\hline
&\textbf{Engineering} & \textbf{Vetting} & \textbf{Hybrid} \\
\hline
\textbf{1. How do interoperation and economic incentives clash (or not)? } & Network effects are more important than interoperation and Apple can use its green bubbles to push iPhone uptake  & Forcing third parties through existing vetting means collecting fees opening that up could result in a loss of revenue & Preventing interoperation keeps other developers from offering products with those functionalities \\ 
\hline
\textbf{2. How do security arguments and economic incentives align (or not)?} & Interoperation could reduce uptake pressure and would be an engineering challenge & Being gatekeeper on security grounds coincides with extracting rent and keeping out competition & If only in-house products can be trusted with certain features, no competitors can offer similar products  \\
\hline
\textbf{3. What needs to change in order for the company to
interoperate? } & Building a system that allows for interoperation and/or moving to a standard protocol & Continuing existing vetting and developing vetting for newly opened avenues, changing pricing schemes & Building in interoperation functionality as well as vetting potential new comers  \\

\hline
\textbf{4. What tension exists between security and other goals
both before/after?} & No tension if you do it right! Getting the whole community involved might make a more secure product & Already allowing some less than perfect actors on the platform; new system could see a loss of revenue and less security if standards are lessened & Already using a vulnerable system; interoperation could improve security or keep it the same if engineering is done well  \\
\hline
\end{tabularx}

\caption{Summary of analysis results}
\label{tab:big_table}
\end{table*}

\begin{table}[ht]
\centering
\renewcommand{\arraystretch}{1.5}
\begin{tabular}{|p{1.7cm}|p{6cm}|p{4.6cm}|p{2.7cm}|}
\hline
\textbf{Policy Area} & \textbf{Legal Case} & \textbf{Technology} & \textbf{SvI Concern Type} \\
\hline
DMA                 & DMA.100097 Meta - Messenger                              & E2EE Messaging            &  Engineering             \\ 
\hline
\multirow{3}{*}{DMA} & \multirow{3}{*}{DMA.100206 Apple - new business terms}  &  App Distribution and IAP & \multirow{3}{*}{Vetting}  \\ 
                                                                    \cline{3-3}
                     &                                              &  Alternative App Stores   &                           \\
                                                                    \cline{3-3}
                     &                                              &  Webkit                   &                           \\
\hline
DMA                  & DMA.100204 Apple -  Physical Devices          & \multirow{3}{*}{OS and Physcial Devices}& \multirow{5}{*}{Hybrid}   \\
\cline{1-2}
DMA                  & DMA.100204 Apple - Process                    &                                          & \\
\cline{1-2}
Merger               & M.9660 Google / Fitbit                    &                                           & \\
\cline{1-3}
Antitrust            & AT.40452 Apple - Mobile payments            &  \multirow{2}{*}{NFC}                     & \\
\cline{1-2}
Merger               & M.8306 Qualcomm / NXP Semiconductors     &                                            &  \\
\hline
Antitrust            & AT.37792 Microsoft                       & Windows OS                                  & \multirow{2}{*}{No - Degrade} \\
\cline{1-3}
Merger               & M.8314 Broadcom / Brocade                & Fiber Channel SAN Switch                    & \\
\hline
Merger               & M.3998 Axalto / Gemplus            & SIM chips                          & \multirow{2}{*}{No - Disclose} \\
\cline{1-3}
Merger               & M.5669 Cisco / Tandberg           & Telepresence Interoperability Protocol      &  \\
\hline
Merger               & M.5984 Intel / McAfee                & Malware detection                  & \multirow{3}{*}{No - Both}\\
\cline{1-3}
Merger               & M.6564 ARM / Giesecke \& Devrient / Gemalto / JV    & Trusted Execution Environment         & \\
\cline{1-3}
Merger               & M.10806  Broadcom / VMware                        & Hardware + Software for virtualization &  \\
\hline
\end{tabular}
\caption{Methodology Results}
\label{tab:all_cases}
\end{table}

\end{document}